# Mean of Ratios or Ratio of Means: statistical uncertainty applied to estimate Multiperiod Probability of Default

**Matteo Formenti**[1]

*Group Risk Management*
*UniCredit Group*

*Università Carlo Cattaneo*

September 3, 2014

**Abstract**

The estimate of a Multiperiod probability of default applied to residential mortgages can be obtained using the mean of the observed default, so called the Mean of ratios estimator, or aggregating the default and the issued mortgages and computing the ratio of their sum, that is the Ratio of means. This work studies the statistical properties of the two estimators with the result that the Ratio of means has a lower statistical uncertainty. The application on a private residential mortgage portfolio leads to a lower probability of default on the overall portfolio by eleven basis points.

---

# The estimates of a Multiperiod probability of default

Consider a class of rating and $N_j$ mortgages issued by the bank in $j = \{1, \dots, T\}$ years. At every issued year it is possible to observe the default at $t$-years of horizon. Table 1 shows an example of mortgages issued in 2006-2009, and the corresponding observation of default from one to the fifth year of time horizon.

|                 | Issued Year |       |       |       |
| --------------- | ----------- | ----- | ----- | ----- |
| Time to default | 2006        | 2007  | 2008  | 2009  |
| def @1Y         | 0           | 0     | 1     | 0     |
| def @2Y         | 1           | 2     | 4     | 2     |
| def @3Y         | 2           | 4     | 6     |       |
| def @4Y         | 2           | 7     |       |       |
| def @5Y         | 3           |       |       |       |
| Issued (N)      | 3.385       | 4.375 | 3.518 | 2.486 |

Table 1 Mortgage issued in Years 2006-2009 and defaults

The real default rate, that is the observed probability of default at $t$-horizon, equals the number of default over the issued mortgages as shown in Equation 1:

$$p_j(t) = \frac{d_j(t)}{N_j}$$

Equation 1 Default rate at t-horizon for mortgages issued in j year

Table 2 applies the default rate at every issued year and for every time horizon given data in Table 1.

|                 | Issued Year |         |         |         |
| --------------- | ----------- | ------- | ------- | ------- |
| Time to default | 2006        | 2007    | 2008    | 2009    |
| def @1Y         | 0,0000%     | 0,0000% | 0,0284% | 0,0000% |
| def @2Y         | 0,0295%     | 0,0457% | 0,1137% | 0,0805% |
| def @3Y         | 0,0591%     | 0,0914% | 0,1706% |         |
| def @4Y         | 0,0591%     | 0,1600% |         |         |
| def @5Y         | 0,0886%     |         |         |         |

Table 2 Default rate

There are two estimators to aggregate information regarding the probability of default at time horizon level:

1. **Mean of Ratios**: it calculates the mean of the each default rate at every time horizon:

$$\mu_{MR}(t) = \mathrm{E}\left[\frac{d_j(t)}{N_j}\right]$$

Equation 2 Mean of ratios

2. **Ratio of Means**: it calculates the ratio between the sum of defaults over the sum of the issued mortgages that is, actually, the ratio of the corresponding means:

$$\mu_{RM}(t) = \frac{\sum_j^T d_j(t)}{\sum_j^T N_j(t)} = \frac{\mathrm{E}[d_i]}{\mathrm{E}[N_i]}$$

Equation 3 Ratio of means

The first estimator is called **Mean of ratios** because each number of the mean is a ratio, while the second one as **Ratio of means** because both the numerator and denominator of the ratio are mean, indeed, if divided by the same number of observation $T$.

(Cochrane, 1977) shows that both estimators are bias when computing the ratio of two volumes by using an application of a Taylor expansion. In this work we study the statistical properties of the two estimators for the purposes of understanding which estimator has the lowest statistical uncertainty when aggregating the default rate over years. The goal is to have a unique probability of default representing the Multiperiod probability of default and the estimator with lowest statistical uncertainty.

In Table 3 we compute the Mean of ratios and the Ratio of means for the example shown in Table 1. The differences are not negligible because there are few number of issued years and the number of mortgages changes sharply. In fact, the two estimators differs in terms of the relative weights that assign to the number of issued years respect to the issued mortgages. In the Mean of ratios estimator the mean assigns a higher weight to the number of issued years because it gives an equally weighted probability to each ratio. On the contrary, the Ratio of means shrinks the volatility of default events therefore it is and estimator less dependent on the number of issued years.

|                 | Issued Year |      |      |      | Mean of Ratios | Ratio of Means |
|-----------------|------|------|------|------|----------------|----------------|
| Time to default | 2006 | 2007 | 2008 | 2009 |                |                |
| def @1Y         | 0    | 0    | 1    | 0    | 0,007106%      | 0,007265%      |
| def @2Y         | 1    | 2    | 4    | 2    | 0,067352%      | 0,065388%      |
| def @3Y         | 2    | 4    | 6    |      | 0,107021%      | 0,106402%      |
| def @4Y         | 2    | 7    |      |      | 0,109542%      | 0,115979%      |
| def @5Y         | 3    |      |      |      | 0,088626%      | 0,088626%      |
| Issued (N)      | 3.385 | 4.375 | 3.518 | 2.486 |              |                |

Table 3 Ratio of means and Mean of ratios

Table 4 shows an example of the potential bias of using the Mean of ratio with respect to the Ratio of mean: a slightly increase in the default (from one to two) leads to an increase of 2.5% in the Mean of ratio estimator, while the same change in the number of issued mortgages lead to a -0.227% difference with respect to the Ratio of mean. This is the case when two dataset are merged such as after the acquisition of a bank and its credit portfolio

In the next section we provide evidences of the statistical properties in order to confirm the result of this examples.

| Year                 | t1  | t2     | t3     | t4     | sum   | mean  | Ratio of means |
|----------------------|-----|--------|--------|--------|-------|-------|----------------|
| Prepayment event (d) | 1   | 10     | 100    | 1000   | 1111  | 277,8 | 10,000%        |
| Number mortgages (N) | 10  | 100    | 1000   | 10000  | 11110 | 2778  | Mean of ratios |
| Hazard ratio (d/N)   | 10% | 10,00% | 10,00% | 10,00% |       |       | 10,000%        |

| Year                 | t1  | t2     | t3     | t4     | sum   | mean  | Ratio of means |
|----------------------|-----|--------|--------|--------|-------|-------|----------------|
| Prepayment event (d) | 2   | 10     | 100    | 1000   | 1112  | 278   | 10,009%        |
| Number mortgages (N) | 10  | 100    | 1000   | 10000  | 11110 | 2778  | Mean of ratios |
| Hazard ratio (d/N)   | 20% | 10,00% | 10,00% | 10,00% |       |       | 12,500%        |

| Year                 | t1  | t2     | t3     | t4     | sum   | mean  | Ratio of means |
|----------------------|-----|--------|--------|--------|-------|-------|----------------|
| Prepayment event (d) | 1   | 10     | 100    | 1000   | 1111  | 277,8 | 9,999%         |
| Number mortgages (N) | 11  | 100    | 1000   | 10000  | 11111 | 2778  | Mean of ratios |
| Hazard ratio (d/N)   | 9%  | 10,00% | 10,00% | 10,00% |       |       | 9,773%         |

Table 4 Differences of the estimators

## Statistical properties

Literature debates about the statistical properties of the Ratio of means and Mean of ratios estimator since (Cochrane, 1977) theoretical result. However, the theoretical result was also confirmed not only when the ratio is between volumes but stereology such as in (Hamdan, Szarka, & Jacks, 2006), or subsample of mineral extraction (Rao, 2002) or a normal population (Rao C. , 1952). According to (Hamdan, Szarka, & Jacks, 2006) the Ratio Of Means is the estimator with the smallest mean square error. This is proven by simulating perfect object (or known values) with noise or a generic object (unknown values with noise). However both the estimators are bias with respect to the Conditional Best Linea Unbiased (CBLU) estimator proposed by

(Baddeley & Jensen, 1991) that assumes each elements of denominator (i.e. mortgages) are conditionally uncorrelated and the variance of denominator conditioned to the numerator needs to be calibrated. As a consequence, the CBLU estimator cannot be applied to the estimation of the probability of default due to the commercial policy that lead to correlation of mortgages issued over time.

On the other hand, the use of two tests were debated from the empirical side such as in (Larivière & Gingras, 2011), that applied the estimators to the citations of research paper, or in the medicine cost effectiveness (Stinnett & Paltiel, 1996).

We provide statistical evidences of the properties of the estimators with a Montecarlo simulation of mortgages and corresponding defaults by using a constant probability of default[2], but slightly perturbed by a Gaussian random variable $\varepsilon \sim N(0,1)$ in order to verify the efficiency of the two estimators:

$$d_i(t) = (p + \sigma \epsilon) N_j$$

**Equation 4 Default rate simulation**

where we control the effect of the perturbation by $\sigma$ and we round the default to integer values. Last we allow the number of issued mortgagee being between 500 and 10.000 according to the real mortgages we have in Table 5-Table 8. Finally, we consider the best estimator as the one with lower relative means square error (RMSE).

Figure 1 shows one simulation of the estimated mean of the two estimators using a perturbation equals to $\sigma = 0.1\%$ and $T = 5, 10$ or $T = 15$. The estimator Mean of ratios has a higher RMSE ($\sigma_{\mu_{MR}} = 3.4798\%$ for $T = 2$, and $\sigma_{\mu_{MR}} = 0.3229\%$ for $T = 10$) with respect to the Ratio of means ($\sigma_{\mu_{MR}} = 0.3479\%$ for $T = 2$, and $\sigma_{\mu_{MR}} = 1.2626\%$ for $T = 10$). We verified the difference of the two RMSE when $T = 1500$, as a robustness check,. As it was expected by increasing the number of issued years the difference is negligible (0.02%) as shown in Figure 2.

---

[2] Probability of default is constant for each time horizon and issued mortgages. We simulate 100 thousands scenarios for each time horizon and issued number of years.

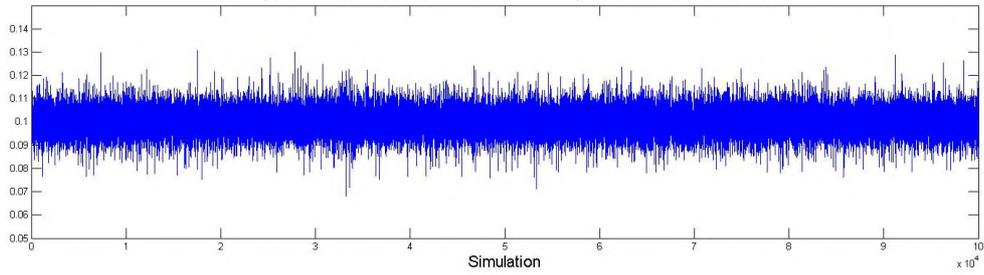

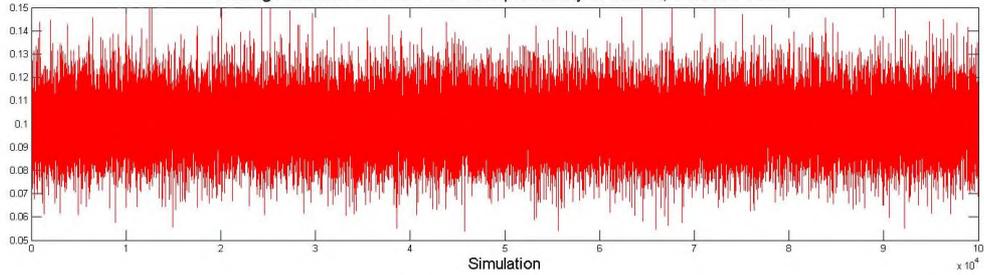

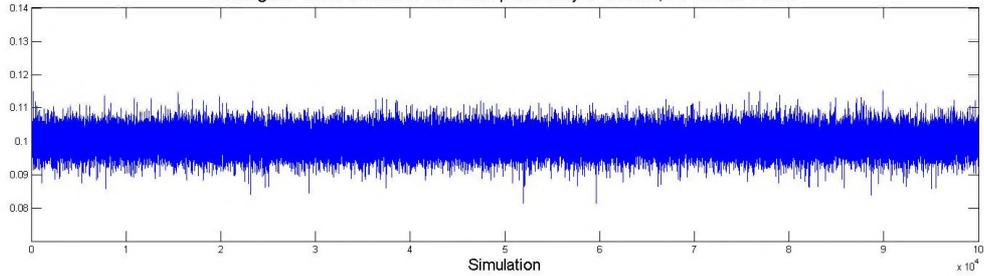

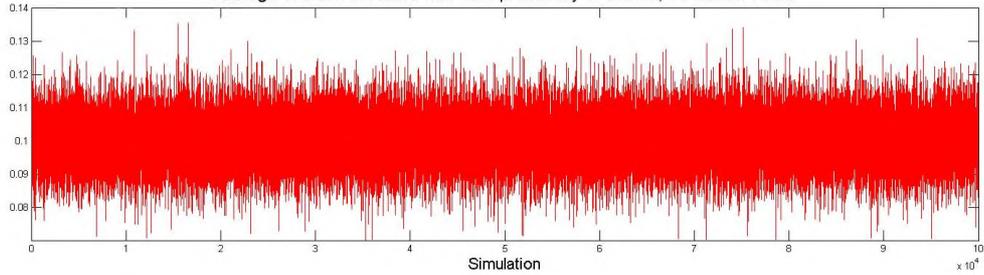

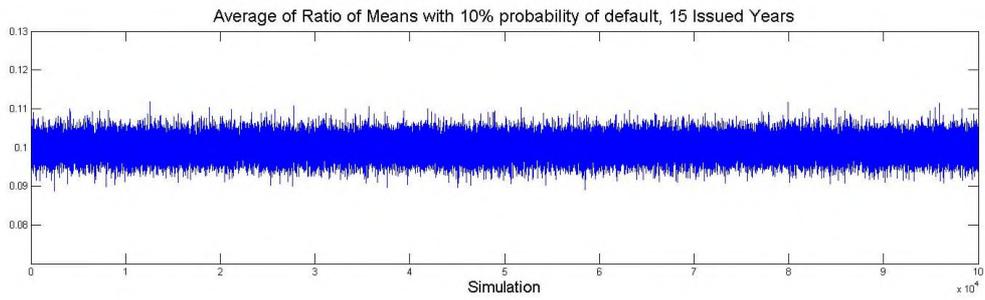
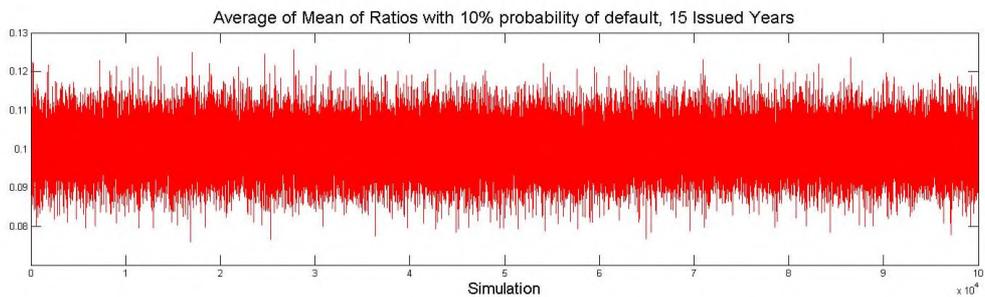

**Figure 1** Average of estimators for T=5,10,15 years

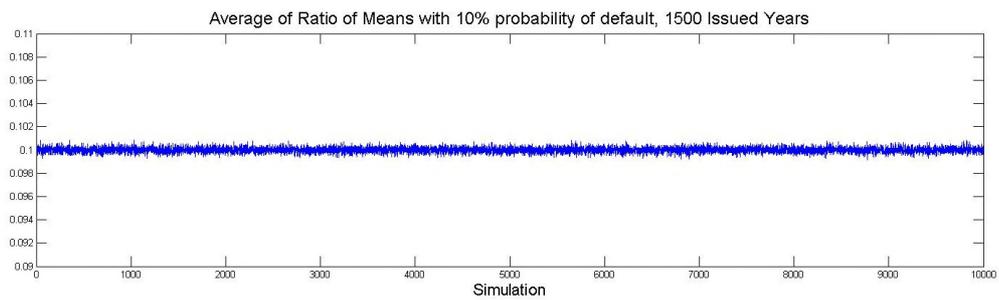
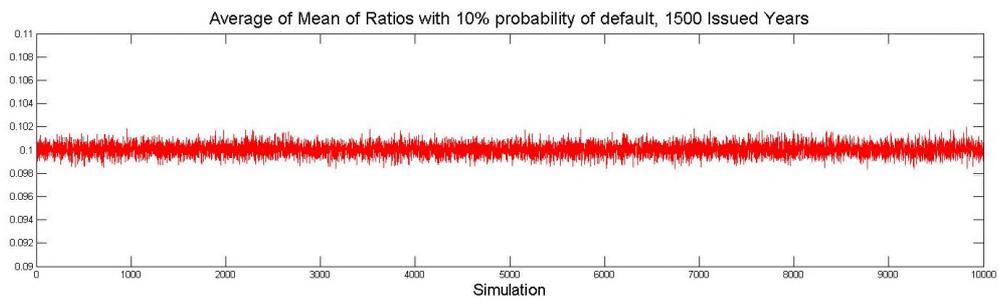

**Figure 2** Average of estimators for T=1500 years

We verify the efficiency by computing the ratio of the two RMSE when the perturbation is low ($\sigma = 0.1$) and a high number of issued years (100 number of years). Figure 3 confirms the results according to the RMSE are decreasing in the number of years, as we expect for any estimator, and the difference of the two is going to stabilize.

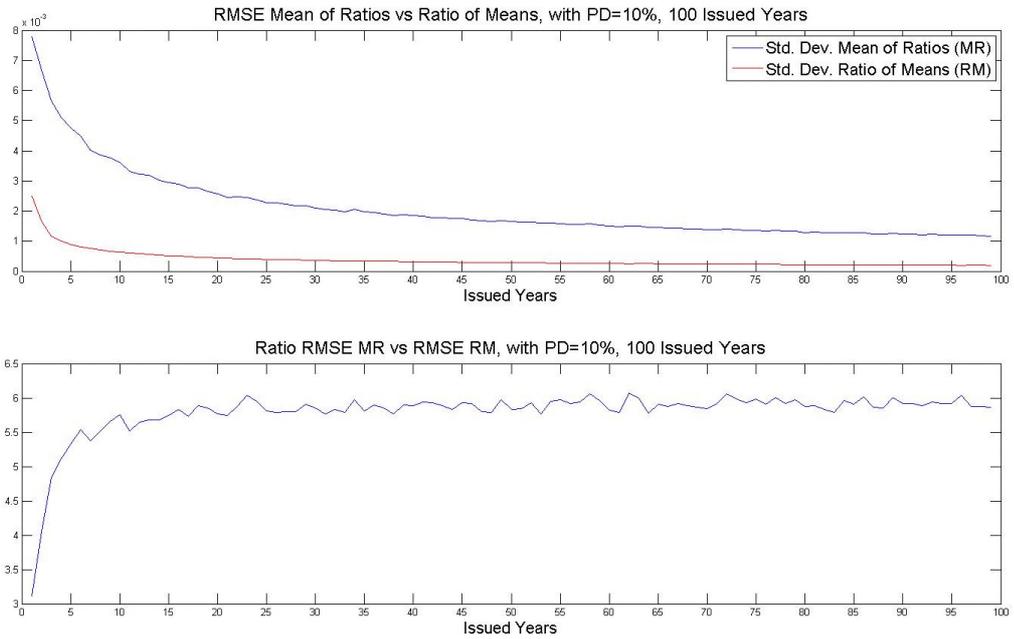

**Figure 3 RMSE Mean of Ratio (MR) and Ratio of Mean (RM) with 100 issued years**

Finally, we compute the ratio of the RMSE using different perturbation $\sigma \in (0,1)$ and a fixed number of issued years $T = 10$, as a robustness check. Figure 4 shows that increasing the perturbation let the Mean of ratios estimator being more uncertain.

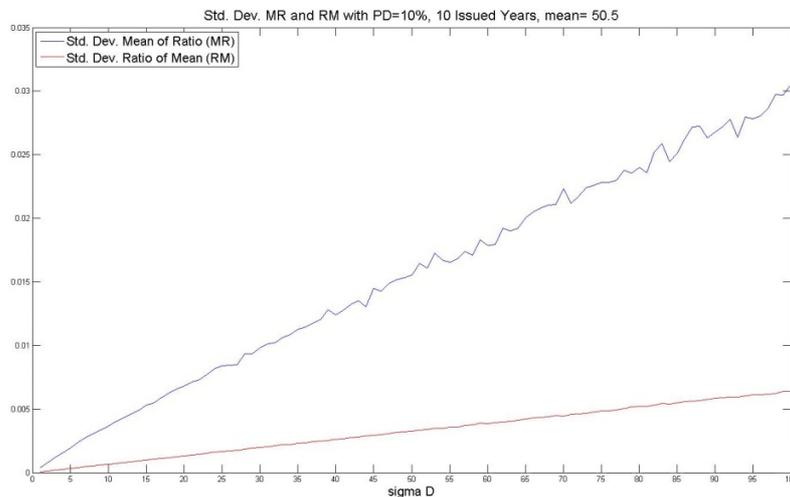

**Figure 4 Standard deviation of MR and RM with different perturbation of the error**

We conclude the Ratio of means is the estimator with the lower relative mean square error, therefore the one with lowest statistical error, that better fit the purposes of aggregating the default rate in order to have a Multiperiod probability of default.

## Empirical evidence using a private mortgage portfolio

We use a portfolio of residential mortgage in the period 2008-2011 in order to understand the difference of the application of the two estimators[3]. The portfolio is splitted at every year in twelve classes of rating. Each estimator is applied in order to have a unique Multiperiod observed probability of default. In Table 5-Table 8 we show the mortgages issued at each year and for each class of rating, and the corresponding default rate at different time horizons. It is to remark that a default rate at five year horizon can be observed only for mortgage issued at least five years ago so there is one observation of default rate, that is the mortgages issued in 2008 and defaulted in 2013 (the fifth year). Finally note that the two estimators can be applied to aggregate the information for each class of rating where we observe a systematic higher difference of the Mean of Ratio with respect the Ratio of means.

| Default 2008 | | | | | | | Default rates | | | | | |
|---|---|---|---|---|---|---|---|---|---|---|---|---|
| Rating | Issued | def @1Y | def @2Y | def @3Y | def @4Y | def @5Y | Rating | def @1Y | def @2Y | def @3Y | def @4Y | def @5Y |
| M01 | 3385 | 0 | 1 | 2 | 2 | 3 | M01 | 0,00% | 0,03% | 0,06% | 0,06% | 0,09% |
| M02 | 3591 | 0 | 0 | 5 | 8 | 15 | M02 | 0,00% | 0,00% | 0,14% | 0,22% | 0,42% |
| M03 | 5243 | 1 | 3 | 19 | 36 | 56 | M03 | 0,02% | 0,06% | 0,36% | 0,69% | 1,07% |
| M04 | 8064 | 1 | 16 | 61 | 100 | 128 | M04 | 0,01% | 0,20% | 0,76% | 1,24% | 1,59% |
| M05 | 7273 | 5 | 31 | 93 | 146 | 206 | M05 | 0,07% | 0,43% | 1,28% | 2,01% | 2,83% |
| M06 | 8696 | 10 | 60 | 168 | 289 | 388 | M06 | 0,11% | 0,69% | 1,93% | 3,32% | 4,46% |
| M07 | 10539 | 28 | 152 | 344 | 478 | 607 | M07 | 0,27% | 1,44% | 3,26% | 4,54% | 5,76% |
| M08 | 7960 | 67 | 223 | 464 | 640 | 790 | M08 | 0,84% | 2,80% | 5,83% | 8,04% | 9,92% |
| M09 | 6149 | 103 | 369 | 718 | 945 | 1124 | M09 | 1,68% | 6,00% | 11,68% | 15,37% | 18,28% |
| M10 | 6558 | 260 | 767 | 1408 | 1867 | 2173 | M10 | 3,96% | 11,70% | 21,47% | 28,47% | 33,14% |
| M11 | 3442 | 294 | 870 | 1387 | 1722 | 1960 | M11 | 8,54% | 25,28% | 40,30% | 50,03% | 56,94% |
| M12 | 775 | 98 | 304 | 465 | 531 | 573 | M12 | 12,65% | 39,23% | 60,00% | 68,52% | 73,94% |
| Total | 71.675 | 867 | 2.796 | 5.134 | 6.764 | 8.023 | Ratio of Means | 1,21% | 3,90% | 7,16% | 9,44% | 11,19% |
| | | | | | | | Mean of Ratio | 2,35% | 7,32% | 12,26% | 15,21% | 17,37% |

Table 5 Mortgage portfolio in 2008

| Default 2009 | | | | | | | Default rates | | | | | |
|---|---|---|---|---|---|---|---|---|---|---|---|---|
| Rating | Issued | def @12 | def @24 | def @36 | def @48 | def @60 | 2007 | def @1Y | def @2Y | def @3Y | def @4Y | def @5Y |
| M01 | 4378 | 0 | 2 | 4 | 7 | | M01 | 0,00% | 0,05% | 0,09% | 0,16% | |
| M02 | 4321 | 1 | 4 | 10 | 23 | | M02 | 0,02% | 0,09% | 0,23% | 0,53% | |
| M03 | 5792 | 2 | 8 | 19 | 30 | | M03 | 0,03% | 0,14% | 0,33% | 0,52% | |
| M04 | 8217 | 4 | 23 | 55 | 93 | | M04 | 0,05% | 0,28% | 0,67% | 1,13% | |
| M05 | 7452 | 6 | 42 | 92 | 152 | | M05 | 0,08% | 0,56% | 1,23% | 2,04% | |
| M06 | 9458 | 17 | 97 | 193 | 288 | | M06 | 0,18% | 1,03% | 2,04% | 3,05% | |
| M07 | 11308 | 46 | 178 | 334 | 513 | | M07 | 0,41% | 1,57% | 2,95% | 4,54% | |
| M08 | 9044 | 87 | 273 | 483 | 663 | | M08 | 0,96% | 3,02% | 5,34% | 7,33% | |
| M09 | 6184 | 127 | 426 | 690 | 936 | | M09 | 2,05% | 6,89% | 11,16% | 15,14% | |
| M10 | 6108 | 325 | 909 | 1392 | 1769 | | M10 | 5,32% | 14,88% | 22,79% | 28,96% | |
| M11 | 3236 | 344 | 991 | 1406 | 1716 | | M11 | 10,63% | 30,62% | 43,45% | 53,03% | |
| M12 | 812 | 152 | 381 | 501 | 574 | | M12 | 18,72% | 46,92% | 61,70% | 70,69% | |
| Total | 76.310 | 1.111 | 3.334 | 5.179 | 6.764 | | Ratio of Means | 1,46% | 4,37% | 6,79% | 8,86% | |
| | | | | | | | Mean of Ratio | 3,20% | 8,84% | 12,67% | 15,59% | |

Table 6 Mortgage portfolio in 2009

---
[3] In this example we do not consider the employee mortgages.

| Default 2010 | | | | | | | Default rates | | | | | |
|---|---|---|---|---|---|---|---|---|---|---|---|---|
| Rating | Issued | def @12 | def @24 | def @36 | def @48 | def @60 | 2008 | def @1Y | def @2Y | def @3Y | def @4Y | def @5Y |
| M01 | 3518 | 1 | 4 | 6 | | | M01 | 0,03% | 0,11% | 0,17% | | |
| M02 | 3291 | 0 | 2 | 7 | | | M02 | 0,00% | 0,06% | 0,21% | | |
| M03 | 4149 | 0 | 6 | 16 | | | M03 | 0,00% | 0,14% | 0,39% | | |
| M04 | 6149 | 10 | 19 | 43 | | | M04 | 0,16% | 0,31% | 0,70% | | |
| M05 | 6163 | 10 | 42 | 87 | | | M05 | 0,16% | 0,68% | 1,41% | | |
| M06 | 7704 | 18 | 72 | 163 | | | M06 | 0,23% | 0,93% | 2,12% | | |
| M07 | 8663 | 44 | 148 | 321 | | | M07 | 0,51% | 1,71% | 3,71% | | |
| M08 | 6484 | 116 | 274 | 459 | | | M08 | 1,79% | 4,23% | 7,08% | | |
| M09 | 4672 | 207 | 425 | 676 | | | M09 | 4,43% | 9,10% | 14,47% | | |
| M10 | 4127 | 405 | 782 | 1109 | | | M10 | 9,81% | 18,95% | 26,87% | | |
| M11 | 1520 | 225 | 459 | 698 | | | M11 | 14,80% | 30,20% | 45,92% | | |
| M12 | 534 | 150 | 283 | 355 | | | M12 | 28,09% | 53,00% | 66,48% | | |
| Total | 56.974 | 1.186 | 2.516 | 3.940 | | | Ratio of Means | 2,08% | 4,42% | 6,92% | | |
| | | | | | | | Mean of Ratio | 5,00% | 9,95% | 14,13% | | |

Table 7 Mortgage portfolio in 2010

| Default 2011 | | | | | | | Default rates | | | | | |
|---|---|---|---|---|---|---|---|---|---|---|---|---|
| Rating | Issued | def @12 | def @24 | def @36 | def @48 | def @60 | 2009 | def @1Y | def @2Y | def @3Y | def @4Y | def @5Y |
| M01 | 2486 | 0 | 2 | | | | M01 | 0,00% | 0,08% | | | |
| M02 | 2072 | 0 | 0 | | | | M02 | 0,00% | 0,00% | | | |
| M03 | 2348 | 0 | 1 | | | | M03 | 0,00% | 0,04% | | | |
| M04 | 3037 | 9 | 21 | | | | M04 | 0,30% | 0,69% | | | |
| M05 | 2740 | 8 | 24 | | | | M05 | 0,29% | 0,88% | | | |
| M06 | 3039 | 17 | 38 | | | | M06 | 0,56% | 1,25% | | | |
| M07 | 3119 | 41 | 110 | | | | M07 | 1,31% | 3,53% | | | |
| M08 | 2287 | 77 | 146 | | | | M08 | 3,37% | 6,38% | | | |
| M09 | 1893 | 102 | 225 | | | | M09 | 5,39% | 11,89% | | | |
| M10 | 1427 | 126 | 228 | | | | M10 | 8,83% | 15,98% | | | |
| M11 | 349 | 40 | 103 | | | | M11 | 11,46% | 29,51% | | | |
| M12 | 108 | 27 | 55 | | | | M12 | 25,00% | 50,93% | | | |
| Total | 24.905 | 447 | 953 | | | | Ratio of Means | 1,79% | 3,83% | | | |
| | | | | | | | Mean of Ratio | 4,71% | 10,10% | | | |

Table 8 Mortgage portfolio in 2011

We aggregate these information comparing the two estimators for each class of rating. Table 9 shows the results where we apply the mean of each issued year of the default rate (ratio) for each rating class in order to compute the Mean of ratios (red table); and we sum the defaults and mortgages of each issued year for each rating class in order to get the Ratio of Means. The overall difference of the two curve is 11 basis points more for the Mean of Ratio.

Finally observe that aggregating the information at rating class level, the mean of the total Mean of ratios estimators (i.e., 1.64%) is always lower than the mean of the single aggregated Mean of ratios (i.e., 3.82%). This is in favor of the analysis computed that supports the Ratio of means (1.57%) instead of the Mean of Ratios of means (3.42%), because of the low number of classes, and the intrinsic stability of the estimator.

| Mean of Ratios | | | | | | Ratio of Means | | | | | |
|---|---|---|---|---|---|---|---|---|---|---|---|
| All Years | def @1Y | def @2Y | def @3Y | def @4Y | def @5Y | All Years | def @1Y | def @2Y | def @3Y | def @4Y | def @5Y |
| M01 | 0,01% | 0,07% | 0,11% | 0,11% | 0,09% | M01 | 0,01% | 0,07% | 0,11% | 0,12% | 0,09% |
| M02 | 0,01% | 0,04% | 0,19% | 0,38% | 0,42% | M02 | 0,01% | 0,05% | 0,20% | 0,39% | 0,42% |
| M03 | 0,01% | 0,10% | 0,36% | 0,60% | 1,07% | M03 | 0,02% | 0,10% | 0,36% | 0,60% | 1,07% |
| M04 | 0,13% | 0,37% | 0,71% | 1,19% | 1,59% | M04 | 0,09% | 0,31% | 0,71% | 1,19% | 1,59% |
| M05 | 0,15% | 0,64% | 1,31% | 2,02% | 2,83% | M05 | 0,12% | 0,59% | 1,30% | 2,02% | 2,83% |
| M06 | 0,27% | 0,98% | 2,03% | 3,18% | 4,46% | M06 | 0,21% | 0,92% | 2,03% | 3,18% | 4,46% |
| M07 | 0,62% | 2,06% | 3,31% | 4,54% | 5,76% | M07 | 0,47% | 1,75% | 3,27% | 4,54% | 5,76% |
| M08 | 1,74% | 4,11% | 6,08% | 7,69% | 9,92% | M08 | 1,35% | 3,55% | 5,99% | 7,66% | 9,92% |
| M09 | 3,39% | 8,47% | 12,43% | 15,25% | 18,28% | M09 | 2,85% | 7,65% | 12,26% | 15,25% | 18,28% |
| M10 | 6,98% | 15,38% | 23,71% | 28,72% | 33,14% | M10 | 6,13% | 14,74% | 23,28% | 28,71% | 33,14% |
| M11 | 11,36% | 28,90% | 43,22% | 51,53% | 56,94% | M11 | 10,57% | 28,35% | 42,58% | 51,48% | 56,94% |
| M12 | 21,11% | 47,52% | 62,73% | 69,60% | 73,94% | M12 | 19,16% | 45,90% | 62,28% | 69,63% | 73,94% |
| Mean of Ratios | 1,64% | 4,13% | 6,96% | 9,15% | 11,19% | Ratio of Means | 1,57% | 4,18% | 6,95% | 9,14% | 11,19% |
| Mean of Mean of Ratios | 3,82% | 9,05% | 13,02% | 15,40% | 17,37% | Mean of Ratio of Means | 3,42% | 8,66% | 12,86% | 15,40% | 17,37% |

Table 9 Probability of Default estimation with Mean of Ratios and Ratio of Means

# Conclusion

This work shows the statistical properties of two estimators, the Mean of ratio and the Ratio of means, when applied to estimate a Multiperiod probability of default of a mortgage portfolio. The main results of the work is that the Montecarlo analyses supports the Ratio of means estimator as the one with lower statistical uncertainty and that result is confirmed when applied to a private residential mortgage portfolio, with an estimated Multiperiod probability of default lower of the entire portfolio by eleven basis points.